\documentclass[%
aip, pra,
reprint,
superscriptaddress,
 amsmath,amssymb,
floatfix,
]{revtex4-1}

\usepackage{graphicx}
\usepackage{dcolumn}
\usepackage{bm}
\usepackage{color}

\begin{document}

\preprint{APS/123-QED}

\title{Color center fluorescence and spin manipulation in single crystal, pyramidal diamond tips}

\author{Richard Nelz}
\author{Philipp Fuchs}
\author{Oliver Opaluch}
\author{Selda Sonusen}
\affiliation{Universit\"at des Saarlandes, FR 7.2 Experimentalphysik, 66123 Saarbr\"ucken, Germany}
\author{Natalia Savenko}
\affiliation{Artech Carbon O\"U, J\~{o}e 5, 10151 Tallinn, Estonia}
\author{Vitali Podgursky}
\affiliation{Tallinn University of Technology, Department of Materials Engineering, Ehitajate tee 5, 19086, Tallinn, Estonia}
\author{Elke Neu}
\email{elkeneu@physik.uni-saarland.de}
\affiliation{Universit\"at des Saarlandes, FR 7.2 Experimentalphysik, 66123 Saarbr\"ucken, Germany}%

\date{\today}


\begin{abstract}
We investigate bright fluorescence of nitrogen (NV)- and silicon-vacancy color centers in pyramidal, single crystal diamond tips which are commercially available as atomic force microscope probes. We coherently manipulate NV electronic spin ensembles with $T_2 = 7.7(3)\,\mu$s. Color center lifetimes in different tip heights indicate effective refractive index effects and quenching. Using numerical simulations, we verify enhanced photon rates from emitters close to the pyramid apex rendering them promising as scanning probe sensors.  
\end{abstract}

\maketitle

\begin{figure}[h!]
	\centering
	\includegraphics[width=1\linewidth]{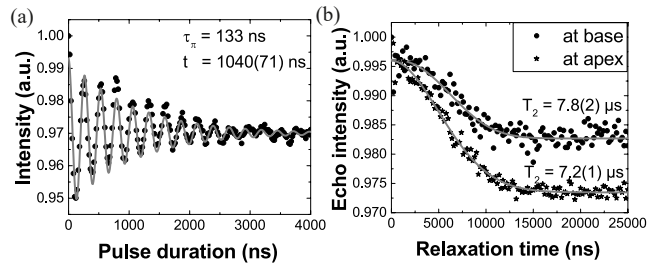}
	\caption{(a) Measurement of a Rabi oscillations between the $m_s=0$ and the $m_s=-1$ states of the Zeeman shifted sub-ensemble with a Rabi frequency of $3.76\,$MHz. (b) Spin echo measurement at apex and base; fitted employing the formalism derived in Ref.\ \onlinecite{Childress2006}}. 
	\label{fig:Pulse}
\end{figure}

Sensitively measuring magnetic fields with high spatial resolution is of paramount importance for many applications in science and industry. Generally, sensing with nanometer resolution requires atomic sized sensors. Point defects in crystals (color centers) localize bound electrons on an atomic scale ($<1\,$nm). Here, color centers in diamond, especially negatively charged nitrogen-vacancy centers (here termed NV centers), are very promising as they are photostable, show bright luminescence and highly-coherent, controllable spins together with optically detected magnetic resonance (ODMR).\cite{Rondin2014} ODMR enables to read out spin states via luminescence, thus even single centers can be used as low back action, quantum-enhanced sensors. Imaging magnetic fields using NV centers e.g.\ contributed insights into superconductivity\cite{Thiel2016} or magnetic materials for spintronics.\cite{Tetienne2015} In life sciences, highly-sensitive measurements of magnetic fields enable nuclear magnetic resonance detection even of single proteins.\cite{Lovchinsky2016} \\
Employing scanning probe methods enables sensing with nanometer resolution (atomic force microscopy AFM).\cite{Rondin2014, Thiel2016, Tetienne2015, Appel2016} In essence, a nanostructure that reliably enables scanning embedded color centers in close proximity to a sample surface (tip-like geometry) is mandatory to fully harness the method's potential.  Optimally, the structure also allows for efficient collection of the color center's fluorescence, consequently enhancing sensitivity. Top-down approaches sculpting scanning probe nanostructures from single crystal diamond\cite{Appel2016} produce stable probes with coherent spins and are thus preferred over approaches where nanodiamonds are attached to tips.\cite{Tetienne2015} However, such approaches require extensive efforts in nanofabrication.\cite{Appel2016} In contrast, bottom-up approaches form nanostructures during diamond growth that are often attached to bulk diamond material and are not straightforwardly usable as scanning probe.\cite{Aharonovich2013} An approach avoiding this problem is the growth of diamond pyramids detachable from the growth substrate and transferable one-by-one to AFM cantilevers.\cite{Obraztsov2010} AFM scanning probes using such pyramids are commercially available\cite{scdonline} and might significantly broaden the application range of scanning probe imaging with color centers.\\
We here demonstrate the basic applicability of commercial single crystal, pyramidal AFM tips for magnetic sensing by showing coherent manipulation of \textit{in-situ} created NV center spins in the devices. We investigate nature and origin of color centers in the tip and simulate how the pyramidal shape influences luminescence collection. Furthermore, we apply reactive ion etching to the tips, illustrating routes towards optimizing ready-to use devices for sensing via removal of diamond.\\
The single crystal diamond AFM tips from Artech Carbon are synthesized in a microwave plasma enhanced chemical vapor deposition (CVD) process based on a CH$_4$/H$_2$ gas mixture. Note that the same CVD technology is used for all tips from Artech Carbon. CVD is performed on silicon (Si) substrates pretreated with micron sized diamond powder to enhance the diamond nucleation density. Refs.\ \onlinecite{Obraztsov2010} and \onlinecite{ObraztsovPatent2009} summarize details on fabrication of geometrically similar diamond tips using an alternative CVD technique. The parameters of our CVD process ensure that (100) facets grow slowest and thus with minimized defect density.\cite{Obraztsov2010} In contrast, on lateral surfaces of crystals growing with (100) facets, highly-defective, nano-crystalline diamond forms. As a consequence, pyramidal diamond micro-crystals with square, (100)-oriented basal planes evolve, embedded into nano-crystalline material. Heat treatment in air entirely removes the latter and transferring the pyramids to AFM cantilevers becomes feasible.\cite{Obraztsov2010} We note that the pyramids are mounted to the cantilevers using epoxy. The apex of the pyramid often has a radius of curvature of $2$ - $20\,$nm, thus rendering these diamonds ideally suitable as ultra-hard AFM tips for high resolution topography imaging or nano-indentation measurements.\\ 
\begin{figure}[h!]
	\centering
	\includegraphics[width=1\linewidth]{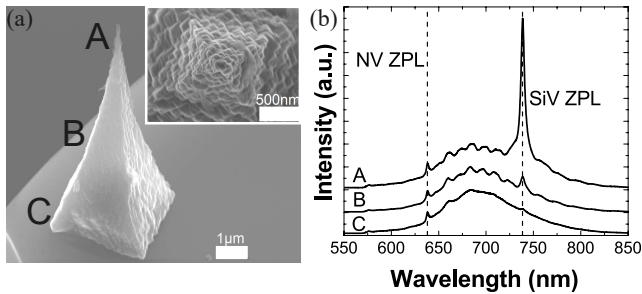}
	\caption{(a) SEM image of a diamond pyramidal tip on a Si cantilever. Inset: View from the apex of the pyramid revealing the step-like surface (b) PL spectra recorded at apex (A), in the middle (B) and at the base of the pyramid (C). The SiV PL is rapidly decreasing compared to the NV PL. Note that all spectra are normalized to maximum of the NV PL. We note that we observe only minor broadband PL due to the epoxy at the pyramid base.}
	\label{fig:SEMSpek}
\end{figure}
To investigate the topography of the pyramids, we record high resolution scanning electron microscopy (SEM) images, as displayed in Fig.\ \ref{fig:SEMSpek}(a). The rough, step-like surface of the pyramid does not straightforwardly reveal its single crystal nature. We prove single crystal nature using NV centers in the tip (see below),  consistent with Raman microscopy characterization provided by the manufacturer.\cite{scdonline} The pyramid in the SEM image has a basal plane of $3.5\times3.5\,\mu\text{m}^2$ and a height of $10\,\mu$m. The cone angle of the tip is as low as $\approx$10$^{\circ}$ close to the apex (last $\approx 0.5\,\mu\text{m}^2$) and 20-25$^{\circ}$ closer to the base.  \\
For sensing and quantum information, negatively charged nitrogen- and silicon-vacancy centers (here termed NV and SiV centers) are of major interest. We investigate the photoluminescence (PL) of both types of color centers in a home built confocal microscope (Numerical aperture NA $0.8$, $532\,$nm laser excitation). Confocal filtering of PL is ensured using a single mode optical fiber. We record PL spectra using a grating spectrometer (Acton Spectra Pro 2500, Pixis 256OE CCD), whereas highly-efficient photon counters (Quantum efficiency $\approx 70\%$, Excelitas SPCM-AQRH-14) enable quantifying PL intensity. To measure PL lifetimes of color centers, we employ pulsed laser excitation (NKT EXW-12, pulse length $\approx 50\,$ps, wavelength $527$ - $537\,$nm) and time correlated photon counting (PicoQuant, PicoHarp 300).\\
In a first step, we investigate the diamond pyramids' PL under laser excitation in general. All 10 investigated pyramids show bright PL with detected photon rates easily exceeding 10$^6$ cps (Mcps) (wavelength range $680$ - $720\,$nm) for moderate excitation power ($<500\,\mu$W) all over the tip. To further quantify the intensity of this PL, which is mainly due to NV centers, we investigate the PL saturation behavior focusing the laser onto the pyramid apex (see supplementary). Estimated NV PL count rates into the first lens of our confocal microscope exceed $500\,$Mcps ($P_{\text{sat}}= 380\,\mu$W) corresponding to an emitted power of $\approx100\,$pW, which is accessible with common commercial photodiodes.\\
We now consider how the pyramids' PL varies with the spatial position of the laser focus. To this end, we record the PL in planes parallel to its basal plane (xy-planes) via scanning the pyramid through the laser focus. To enable 3-dimensional imaging, we change the position of the laser focus with respect to the height of the pyramid (z-position) and repeat the measurement. Focusing the laser close to the basal plane clearly reveals the square footprint of the pyramid as deduced from SEM images, while focusing the laser close to the pyramid's apex reveals PL stemming from a single point (data see supplementary).\\
We now evaluate PL spectra recorded in different heights of the pyramid [see Fig.\ \ref{fig:SEMSpek}(b)]. For all heights, we clearly observe the characteristic spectrum of NV centers with a zero-phonon line (ZPL) at $638\,$nm [linewidth $4.1(2)\,$nm] and broad phonon sidebands between $650$ and $800\,$nm. When exciting PL closer to the pyramid apex, we furthermore observe the ZPL of SiV centers at $738\,$nm [linewidth $4.9(1)\,$nm]. SiV PL is significantly more pronounced the closer to the pyramid apex a spectrum is recorded. For spectra recorded at the apex, the SiV ZPL strongly dominates over the NV phonon sideband. PL spectra with the described features are consistently observed from all 10 investigated tips.\\
\begin{figure}[h!]
	\centering
	\includegraphics[width=1\linewidth]{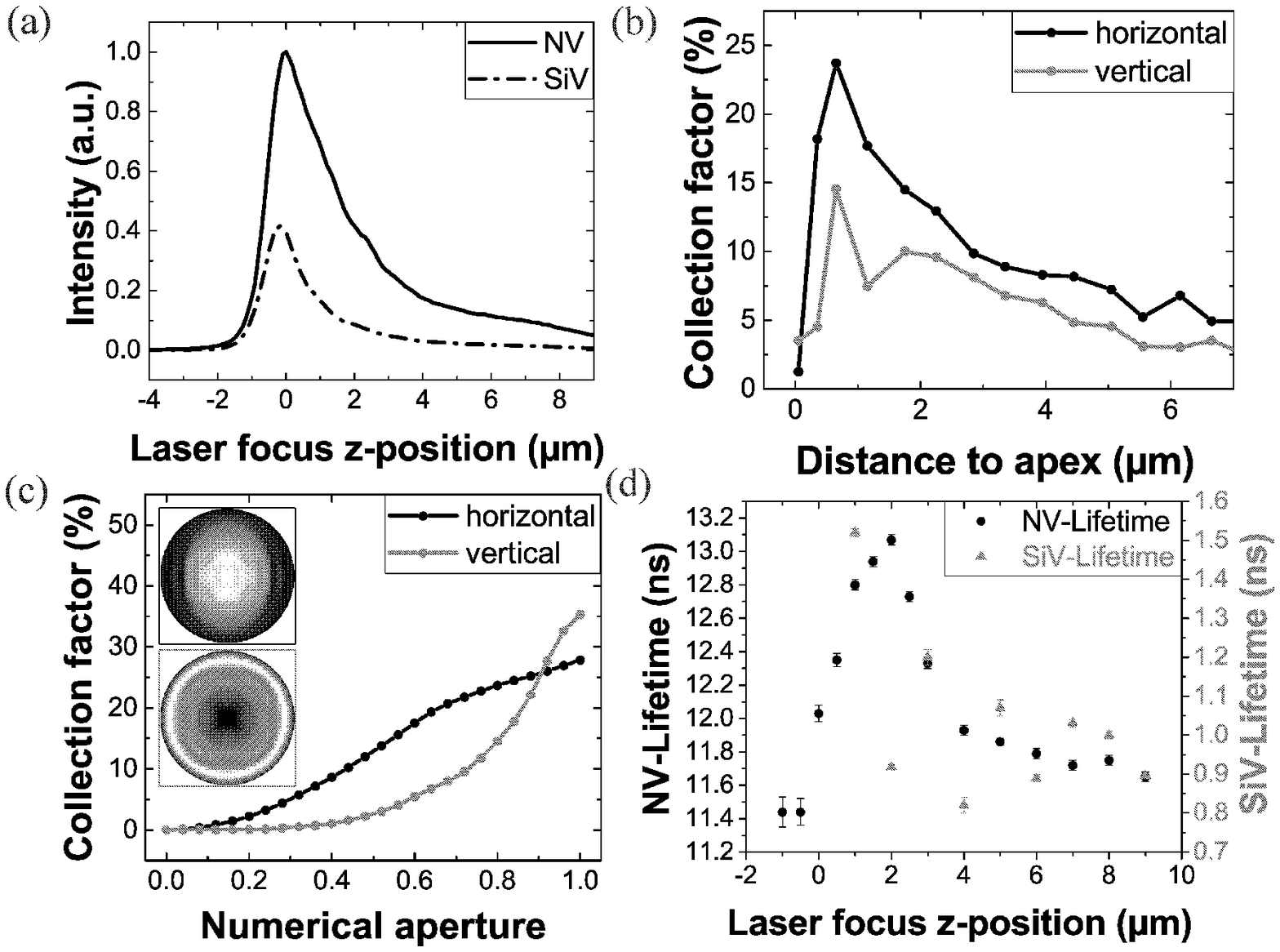}
	\caption{(a) NV and SiV PL along the z-axis of the tip. (b) Simulated collection factor along the z-axis of the tip for horizontal and vertical dipoles. (c) Simulated collection factors for different NA and far field patterns as inset. For these simulations, we place the dipole at the location that maximizes the collection factor [see (b)]. (d) Measured excited state lifetimes along the z-axis of the pyramid; note the increase of the lifetime towards the apex as well as a decrease very close to the apex. The z-axis origin ($z=0$) represents the tip apex in (a) and (d) determined as the z-position for which maximally spatially confined PL from the tip occurs (note the depth of focus of our confocal microscope of $\approx 0.9\,\mu$m).}
	\label{fig:SimuLT}
\end{figure}
To interpret the PL spectra, it is useful to recall that the pyramid's apex is formed in the initial phase of CVD growth on a Si substrate: A CVD plasma etches Si substrates, consequently diamond films grown on Si contain up to $10^{19}\,\text{cm}^{-3}$ Si atoms.\cite{Barjon2005} However, Si incorporation decreases significantly by more than one order of magnitude as soon as the growing diamond fully covers the substrate.\cite{Barjon2005} Thus, we expect a higher concentration of Si impurities close to the apex formed in the starting phase of the growth. In contrast, nitrogen has been added to the gas mixture during the CVD process (0.1\% volume fraction) and forms NV centers in all heights in the diamond pyramid. We investigate the brightness of NV and SiV PL in dependence of the z-position of the laser focus in more detail as shown in Fig.\ \ref{fig:SimuLT}(a). Both PL signals decrease in brightness from apex to base, however, the SiV PL decreases faster, indicating a stronger confinement of the area with high SiV density close to the apex.\\
Despite a significantly reduced diamond volume exposed to the laser, NV and SiV PL is enhanced close to the apex, which we also attribute to photonic effects: we simulate the emission patterns of dipoles with horizontal and vertical alignment with respect to the pyramid base in the middle of the xy-plane for different z-positions. We use the Finite Difference Time Domain (FDTD) method implemented with a commercial software (Lumerical FDTD solutions). As sensing applications often demand to maximize the PL rate in the detection optics, we investigate the collection factor defined as the ratio of the power emitted into the NA to the emitted power in homogeneous diamond as a figure of merit, exemplarily for $\lambda=700$ nm. The results of these simulations are displayed in Fig.\ \ref{fig:SimuLT}(b) and (c) (details see supplementary). The simulated data, that qualitatively agree with our measurements [Fig.\ \ref{fig:SimuLT}(a)], show that the tip's shape leads to an enhanced photon rate in our NA for PL from color centers close to the apex. This situation is advantageous for scanning probe applications, as these centers are simultaneously in close proximity to the sample surface. Moreover, our simulations indicate that (for horizontal dipoles), even a moderate NA of $\approx0.6$ only reduces the collection factor by $\approx25\,\%$.\\
To determine if the diamond tips are suitable for magnetometry applications, we perform ground state optically detected magnetic resonance (ODMR) of the NV centers. Applying a static magnetic field $\vec{B}$ reveals four pairs of ODMR resonances, corresponding to the four equivalent $\left<111\right>$ orientations of NV centers in the diamond lattice (data see supplementary). The occurrence of four pairs of resonances witnesses the single crystal nature of the tip as any misoriented, polycrystalline grains would lead to additional resonances or blurring of the ODMR-spectrum. For the following measurements, we align $\vec{B}$ onto one of the $\left<111\right>$ axes and use the NV sub-ensemble to which $\vec{B}$ is aligned. We verify coherent manipulation of NV electronic spins via Rabi oscillations [see Fig.\ \ref{fig:Pulse}(a)] and measure their coherence times $T_2^*$ and $T_2$. \\
Using a Spin echo-sequence, we measure $T_2$ [see Fig.\ \ref{fig:Pulse}(b)] at the base of 4 pyramids and find on average $T_2 = 7.7(3)\,\mu$s. We check $T_2^*$ via a Ramsey-type experiment and find $T_2^* = 0.24(7)\,\mu$s (data not shown). We verify $T_2 = 7.2(1)\,\mu$s for NVs close to one pyramid's apex [Fig.\ \ref{fig:Pulse}(b)]. Thus, $T_2$ is preserved despite SiV incorporation and potentially lower crystal quality. $T_2$ and $T_2^*$ are almost one order of magnitude higher than in high pressure high temperature diamond, (typ.\ $100\,$ppm substitutional nitrogen $\text{[N]}^\text{s}$, $T_2\approx1\,\mu$s\cite{Rondin2014}) however much lower than in high purity CVD diamond ($1\,$ppb $\text{[N]}^\text{s}$, $T_2\approx300\,\mu$s\cite{Rondin2014}). Using $T_2$ and $T_2^*$ as well as the PL countrate $I_0 = 162.7\,$Mcps, the readout duration $t_L = 350\,$ns and the contrast of the ODMR $C = 7.4\,\%$, we calculate the DC and AC magnetic field sensitivities:\cite{Rondin2014}
\begin{eqnarray}
\eta_{\text{DC}} \approx \frac{1}{\gamma_{\text{NV}}} \frac{1}{C\sqrt{I_0 t_L}} \,\textrm{x}\,  \frac{1}{\sqrt{T_2^*}} \approx  130.6\,\text{nT}/\sqrt{\text{Hz}}   \\
\eta_{\text{AC}} = \eta_{\text{DC}} \,\textrm{x}\, \sqrt{\frac{T_2^*}{T_2}}   \approx 22.9\,\text{nT}/\sqrt{\text{Hz}}
\end{eqnarray}
Previous works report that lifetimes changes of color centers evidence interaction with various defects\cite{Smith2010, Chen2012, Monticone2013} as well as the photonic environment.\cite{Beveratos2001} To investigate lifetime changes in detail, we measure the excited state lifetime $\tau_{\text{NV}}$ ($\tau_{\text{SiV}}$) of NV (SiV) centers along the z-axis of the pyramid [see Fig.\ \ref{fig:SimuLT}(d) and supplementary]. At the base (z $\approx10\,\mu$m), where we identified a reduced density of SiV centers, we find $\tau_{NV} = 11.79(3)\,$ns, which well agrees with NV lifetimes in single crystal bulk diamonds [$\tau_{\text{NV},\text{bulk}}=12.9(1)\,$ns]\cite{Collins1983} whereas $\tau_{\text{SiV}} = 0.89(1)\,$ns, as previously observed in polycrystalline diamond.\cite{Turukhin1996} Approaching the pyramid apex (z $\approx2-4\,\mu$m), $\tau_{\text{NV}}$ [$\tau_{\text{SiV}}$] increase to $13.07(3)\,$ns [$1.92(9)\,$ns], potentially due to a decrease in the effective refractive index close to the nanoscale apex.\cite{Beveratos2001} In the region of increased SiV density very close to the apex (z $\approx0-1\,\mu$m), $\tau_{\text{NV}}$ drops to $11.44(8)\,$ns. In the presence of graphitic or disordered carbon, $\tau_{\text{NV}}$ shortens.\cite{Smith2010, Chen2012} In this context, a high level of Si doping, here possibly induced by substrate etching,\cite{Barjon2005} introduces structural defects.\cite{Bolshakov2015}  Thus, this shortening of $\tau_{\text{NV}}$ might be attributed to incorporated Si potentially lowering crystalline quality and inducing non-diamond phases. Moreover, $\text{[N]}^\text{s}$ can significantly quench NV PL via dipole-dipole interaction.\cite{Monticone2013} Such energy transfer (F\"orster resonance energy transfer, FRET) requires an overlap of the quenched color center's (donor) emission band and the quenching defect's (acceptor) absorption band. Thus, also a high density of SiV centers potentially quenches NV PL: NV emission ($\approx 650$ - $750\,$nm) well overlaps with SiV absorption ($540$ - $740\,$nm\cite{Rogers2014}), especially with its dominant ZPL.  FRET between NV and SiV centers has so far not been demonstrated, however, NV centers show efficient FRET transfer to dye molecules with absorption in a similar spectral range ($50\,\%$ efficiency at $\approx 6\,$nm distance).\cite{Mohan2010} We estimate the average distance between centers in the tip to be $<10\,$nm (see supplementary), thus FRET might be contributing to the quenching. \\
As a first step towards optimization of the tip for magnetometry, we etch it in an inductively-coupled reactive ion etching plasma (Ar/O$_2$, $50\,$sccm each, $18.9\,$mTorr, $500\,$W ICP, $200\,$W RF power see also supplementary) enabling highly-anisotropic etching of diamond. We remove up to $\approx 1\,\mu$m of diamond, as measured from the pyramid height reduction, in several etch steps. We note that the pyramid still reveals a comparably defined apex of \mbox{$<$200 nm} size and NV centers close to the apex retain coherence ($T_2^\textrm{(base, unetched)}= 9.4(2)\,\mu$s; $T_2^\textrm{(apex, etched)}= 8.9(2)\,\mu$s). Surface roughness remains mainly unchanged, while SiV PL significantly decreases, as we remove highly-Si-doped diamond (details see supplementary). We thus establish a procedure to tune NV/SiV ratio even for readily mounted devices to tailor them for magnetometry. \\
In conclusion, we show that single crystal diamond AFM tips from Artech Carbon show bright PL from \textit{in-situ} created NV and SiV centers, easily measurable with standard photodiodes. In addition, we demonstrate their suitability for magnetometry ($T_2 = 7.7(3)\,\mu$s, $\eta_{\text{DC}}\approx  130.6\,\text{nT}/\sqrt{\text{Hz}}$, $\eta_{\text{AC}} \approx 22.9\,\text{nT}/\sqrt{\text{Hz}}$). FDTD simulations indicate high PL rates from centers close to the pyramid apex.\\
The outstanding advantage of the pyramidal tips is their commercial availability as well as a comparably low production cost per tip and high resolution topography imaging, compared to top-down approaches.\cite{Appel2016} 
The main drawback is a high color center density precluding to use single centers for imaging. Consequently, color centers in $\approx1\,\mu$m tip height, limited by the confocal microscope, contribute to sensing and significantly limit spatial resolution. 
Nevertheless, the tips might e.g.\ be utilized to monitor the activity of integrated circuits via magnetic measurements,\cite{Shao2016} here nanoscale resolution is not mandatory and scanning devices avoid the necessity to place a diamond chip on the circuit under investigation.
For single color center sensors, diamond purity has to be enhanced by reducing nitrogen in the process and the use of more etch resistant growth substrates (e.g.\ iridium).  Creation of color centers might be feasible via recently developed nanoimplantation techniques,\cite{RiedrichMoeller2015} whereas mounting of the devices to transparent cantilevers e.g.\ made from silicon nitride would allow optical access through the scanning device.\\
\textbf{Supplementary material}\\
{Please see supplementary material for additional information about our experimental setup, ODMR-spectra, technical details on the plasma treatment, the 3D fluorescence imaging, fluorescence saturation and details on the numerical simulations.\\
This research has been funded via the NanoMatFutur program of the German Ministry of Education and Research (BMBF) under grant number FKZ13N13547 and Enterprise Estonia (EAS) project Lep16012. We thank J\"org Schmauch for recording SEM images as well as Dr.\,Rene Hensel, Dr.\,Tobias Kraus, Dr.\,Daniel Brodoceanu and Susanne Selzer (INM, Saarbr\"ucken) for providing plasma etching tools and assistance.

\end{document}